\documentclass[twocolumn,aps,superscriptaddress,nofootinbib]{revtex4-1}
\usepackage[dvips]{graphicx}
\usepackage{latexsym,amssymb,amsmath}
\usepackage{color}
\usepackage[bookmarksnumbered,bookmarksopen,colorlinks,citecolor=red,linkcolor=blue]{hyperref}
\usepackage{mathrsfs}
\usepackage{times}
\usepackage{csquotes}

\begin{document}

\title{Detecting Chiral Magnetic Effect via Deep Learning}

\author{Yuan-Sheng Zhao}
\email{yczhao19@fudan.edu.cn}
\affiliation{Physics Department and Center for Particle Physics and Field Theory, Fudan University, Shanghai 200433, China}
\author{Lingxiao Wang}
\email{lwang@fias.uni-frankfurt.de}
\affiliation{Frankfurt Institute for Advanced Studies, Ruth Moufang Strasse 1, D-60438,
Frankfurt am Main, Germany}
\author{Kai Zhou}
\email{zhou@fias.uni-frankfurt.de}
\affiliation{Frankfurt Institute for Advanced Studies, Ruth Moufang Strasse 1, D-60438,
Frankfurt am Main, Germany}
\author{Xu-Guang Huang}
\email{huangxuguang@fudan.edu.cn}
\affiliation{Physics Department and Center for Particle Physics and Field Theory, Fudan University, Shanghai 200433, China}
\affiliation{Key Laboratory of Nuclear Physics and Ion-beam Application (MOE), Fudan University, Shanghai 200433, China}
\date{\today}

\begin{abstract}
The search of chiral magnetic effect (CME) in heavy-ion collisions has attracted long-term attentions. Multiple observables have been proposed but all suffer from obstacles due to large background contaminations. In this Letter, we construct an observable-independent \textit{CME-meter} based on a deep convolutional neural network. After being trained over data-set generated by a multiphase transport model, the CME-meter shows high accuracy in recognizing the CME-featured charge separation from the final-state pion spectra. It also exhibits remarkable robustness to diverse conditions including different collision energies, centralities, and elliptic flow backgrounds. In a transfer learning manner, the CME-meter is validated in isobaric collisions, showing good transferability among different collision systems. Moreover, based on variational approaches, we utilize the \textit{DeepDream} method to derive the most responsive CME-spectrum that demonstrates the physical contents the machine learns.
\end{abstract}

\maketitle

%% Introduction

\emph{Introduction}.-- Quantum chromodynamics (QCD) is the standard theory describing the physics of the strong interaction. Among the studies on QCD, the proposal of using chiral magnetic effect (CME) to reveal the vacuum structure of QCD is of great importance~\cite{Kharzeev_1998,Kharzeev:2007jp,fukushima:2008chiral}. It predicts that in hot and dense quark-gluon plasma (QGP), the topological fluctuations of gluon fields can cause imbalance between the number of left-handed and right-handed quarks, and this difference can induce an electric current under external magnetic field.

High energy heavy ion collisions (HICs) provide an environment for CME to take place. However, QGP and strong magnetic field required for giving rise to CME exist only in the early stages of the collisions. To retrieve the information of possible CME from the final-state hadrons, multiple observables were proposed~\cite{Voloshin:2004vk,Deng:2014uja,Magdy:2017yje,Zhao:2017nfq,Tang_2020}, such as the $\gamma$-correlator (see definition in below). However, due to the large contributions of elliptic flow and other background noises~\cite{bzdak:2011azimuthal,Schlichting:2010qia,Wang:2009kd}, these observables can not clearly recognize CME or its induced charge separation (CS) in QGP along the magnetic-field direction. %If only we know how CME evolves exclusively with time in the collision system, then we may not need to suffer the background contamination by constructing a proper observable, or at least have a better understanding on the final states.

Although it is difficult to detect CME through specific observables, analyzing the final-state hadronic spectrum as a whole in the sense of big data would help reveal hidden fingerprints of CME. Deep learning is a branch of machine learning which shows its powerful ability in recognizing patterns and structures with complex correlations~\cite{schmidhuber:2015deep,lecun:2015deep}. With the hierarchical structure of artificial neural networks, deep learning is particularly effective in tackling complex systems that cannot be easily handled by conventional techniques. Recently, significant progress has been made in applying deep learning to physics studies, including nuclear physics~\cite{pang:2018equationofstatemeter,zhou:2019regressive,liu:2019principal,Huang:2018rtk,du:2020identifying,omanakuttan:2020fast,jiang:2021deep, Shi:2021qri}, particle physics~
\cite{Baldi:2014kfa,Baldi:2014pta,Barnard:2017parton,broecker:2017machine,Radovic:2018dip}, and condensed matter physics~\cite{carrasquilla:2017machine,vannieuwenburg:2017learning,Han_2018,carleo:2019machine,wang:2020continuousmixture}. In this Letter, we explore the possibility of using deep learning to determine whether there are detectable final-state signals of CME that survive the collision dynamics and background interference, thus providing us a new feasibility to detect CME in HICs.

%%%%%%%%%%%%%%%%%%%%%%%%%%%%%%%%%%%%%%%%%%%%%%%%%%%%%%%%%%%%%%%%%%%%%%%%%%%%%%%%%%%%%%%%%%%%%%%%%%%%%%%%%
\begin{figure}[htbp!]
\begin{center}
    \includegraphics[width=8.6cm]{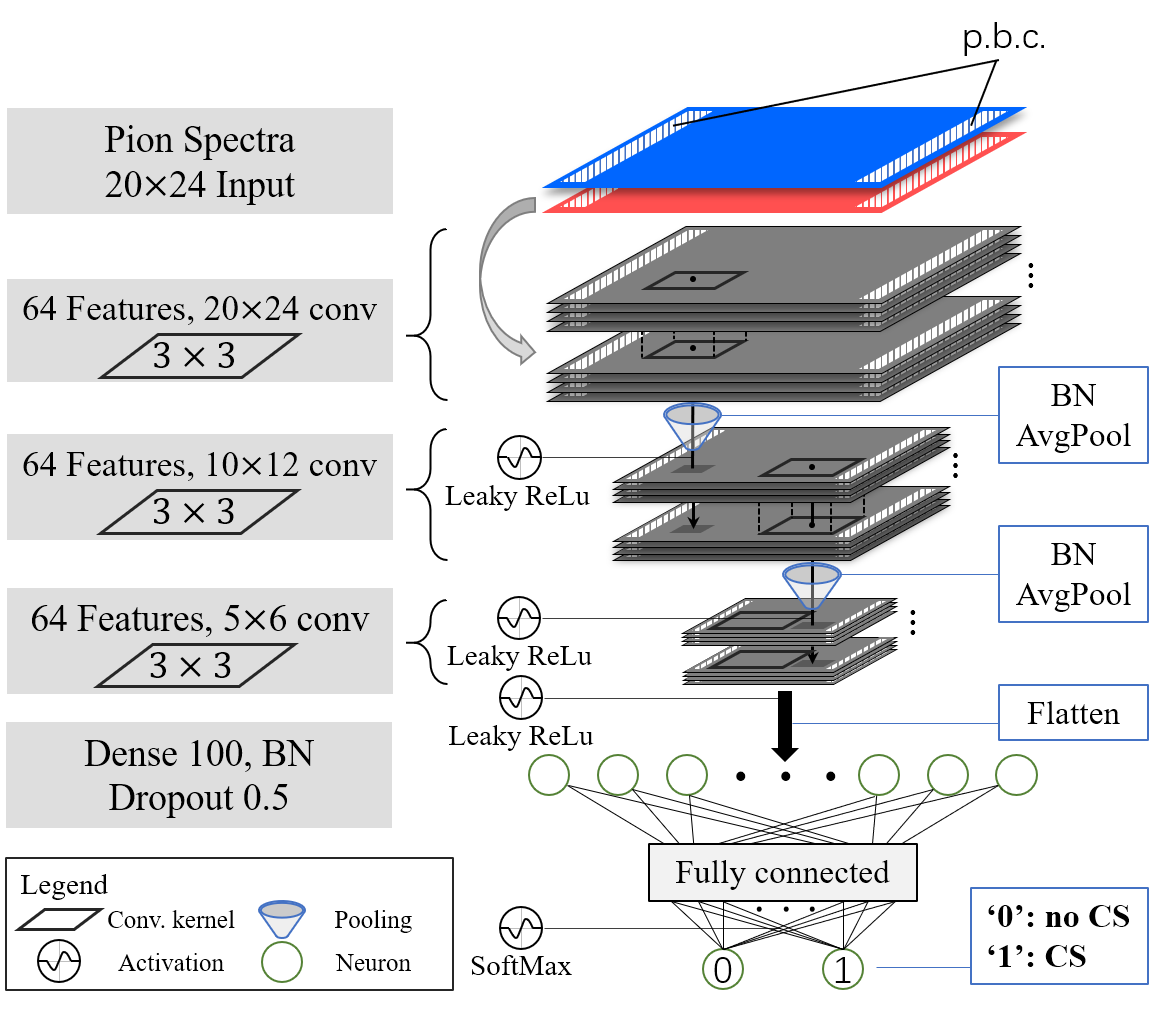}
    \caption{The convolutional neural network architecture with $\pi^+$ and $\pi^-$ spectra $\rho^{\pm}(p_T, \phi)$ as input.
    %where $\phi$ is the angle relative to the reaction plane. Every bin of the spectrum represents a $\Delta p_{T}\times\Delta\phi = 0.2GeV \times \pi/12$ range. In the end we got two spectra with $20\times24$ bins, or two photos of momentum space $\pi^+$ and $\pi^-$ distribution of 480 pixels for each event.
    The periodic boundary condition (\textit{p.b.c}) is recovered both for the input spectra and the 2D convolution operation.}
    \label{fig:CNN}
\end{center}
\end{figure}
%%%%%%%%%%%%%%%%%%%%%%%%%%%%%%%%%%%%%%%%%%%%%%%%%%%%%%%%%%%%%%%%%%%%%%%%%%%%%%%%%%%%%%%%%%%%%%%%%%%%%%%%%
\emph{Method}.-- In this section, we introduce a deep learning model containing convolutional neural networks (CNNs) to detect the CME signals in HIC systems. The architecture of the neural network is shown in Fig.~\ref{fig:CNN}. For the purpose of supervised learning, we prepare the training data-set from the string melting AMPT model~\cite{lin:2005multiphase}~\footnote{The AMPT model is a transport model which is widely used to simulate the evolution of both partonic and hadronic matter in HICs and has been proven to be successful in describing the experimental data of harmonic flows~\cite{Lin:2014tya,Xu:2010du,Bhalerao:2013ina,Wei:2018xpm}, global polarization~\cite{Li:2017slc,Sun:2017xhx,Wei:2018zfb}, QCD phase transitions~\cite{jin:2018explore}, and so on. %consists of four major physical procedures describing the evolution process in HICs:  Initial Condition, Parton-Cascade, Hadronization, and Hadronic Re-scatterings.
}. In order to implement the CME in AMPT model, we adopt a global CS scheme first employed for Au+Au collisions in Ref.~\cite{Ma_2011}. Within such a scheme, the $y$-components of momenta of a fraction of downward moving $u$ quarks and upward moving $\Bar{u}$ quarks are switched, likewise for $\Bar{d}$ and $d$ quarks (\enquote{upward} and \enquote{downward} are refer to the $y$-axis which is perpendicular to the reaction plane). This gives a right-dominant CME event, and for a left-dominant one, we switch quarks with momenta opposite to those in the right-dominant case, or just rotate the event along the $z$-axis (along beam direction) by $180^\circ$. The CS fraction $f$ is introduced as% following, which means the number of quarks to be switched at the beginning of evolution,
\begin{equation}
    f=\frac{N_{\uparrow(\downarrow)}^\pm-N_{\downarrow(\uparrow)}^\mp}{N_{\uparrow(\downarrow)}^\pm+N_{\downarrow(\uparrow)}^\mp},
\end{equation}
where superscript labels the sign of charge and subscript labels the direction of momentum along $y$-axis. Events with $f = 0\%$ belong to \enquote{no CS} class and are labeled as \enquote{0}, while events with $f> 0\%$ are in \enquote{CS} class and labeled as \enquote{1}.

With supervised learning, the deep CNNs are trained to distinguish these two classes from the labeled data. As for the input to the CNNs, we prepare from AMPT simulation a series of 2-D spectra $\rho^{\pm}(p_T, \phi)$ of charged pions ($\pi^+$ or $\pi^-$) in the final state with 20 transverse momentum $p_T$-bins and 24 azimuthal angle $\phi$-bins~\footnote{We maintain the \emph{p.b.c} in $\phi$ in all the convolution layers in the network, so as not to lose correlations near $\phi$ = 0 and 2$\pi$, while $p_T$ follows the original boundary condition of the convolution layer.}. %Since electric charges are mainly carried by pions in such HIC systems, it would be an efficient input addressing the recognition of CS signals.
The data-set consists of Au+Au collisions with colliding energy $\sqrt{s_{NN}}$=7.7, 11.5, 14.5, 19.6, 27, 39, 62.4 and 200 GeV, with CS fraction $f$ = 0\% or $f$ $>$ 0\%, all divided into 6 centrality bins in the range 0$\sim$60\%. Each species of collision conditions contains 50,000 events. To reduce fluctuations, 100 events with the same collision condition and dominant chirality are randomly picked out. Their pion spectra are averaged and normalized to form a single sample in our training batch. To preserve the mirror symmetry rooted in data, every sample is accompanied by its copy which is flipped along $y$-axis. It can also be viewed as exchanging the initial distribution of nucleons between the projectile and target nuclei, which naturally provides data augmentation and reduces redundancy in training the CNNs.

%%%%%%%%%%%%%%%%%%%%%%%%%%%%%%%%%%%%%%%%%%%%%%%%%%%%%%%%%%%%%%%%%%%%%%%%%%%%%%%%%%%
\begin{table}[htbp!]
\caption{The validation accuracy of the well-trained model in confronting unseen events with the same set.}
\label{tab:acc}
\centering
\begin{tabular}{ccc}
\hline
Model             \quad\quad    & 0\%+5\%  \quad\quad & 0\%+10\% \\ \hline
Validation accuracy \quad\quad  & 78.47\%   \quad\quad & 93.38\% \\ \hline
\end{tabular}
\end{table}
%%%%%%%%%%%%%%%%%%%%%%%%%%%%%%%%%%%%%%%%%%%%%%%%%%%%%%%%%%%%%%%%%%%%%%%%%%%%%%%%%%%
%%%%%%%%%%%%%%%%%%%%%%%%%%%%%%%%%%%%%%%%%%%%%%%%%%%%%%%%%%%%%%%%%%%%%%%%%%%%%%%%%%%
\begin{figure}[htbp!]
    \includegraphics[width=7cm]{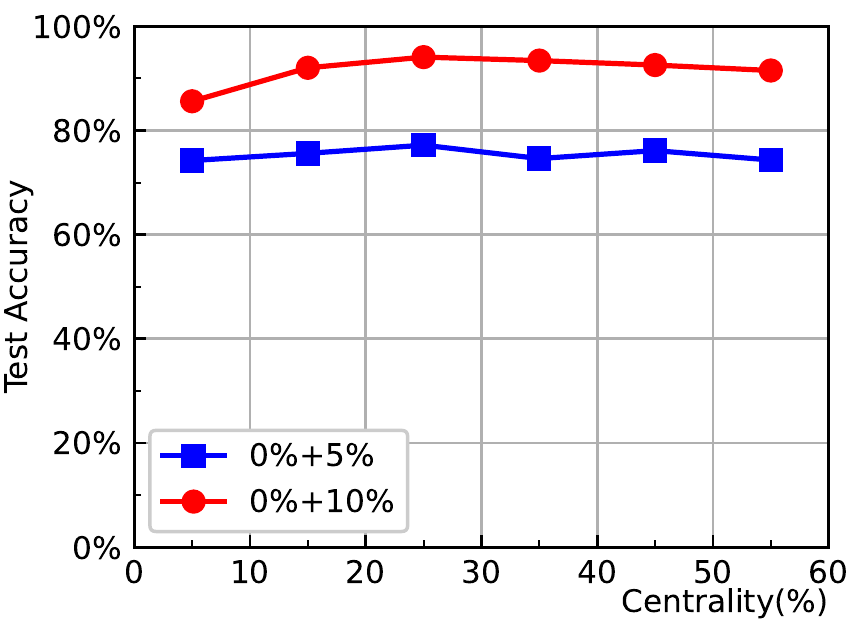}
    \caption{The test accuracy of models on different data-sets containing mixed collision energies along with centralities.}
    \label{fig:Acc_vs_cen}
\end{figure}
%%%%%%%%%%%%%%%%%%%%%%%%%%%%%%%%%%%%%%%%%%%%%%%%%%%%%%%%%%%%%%%%%%%%%%%%%%%%%%%%%%%
%%%%%%%%%%%%%%%%%%%%%%%%%%%%%%%%%%%%%%%%%%%%%%%%%%%%%%%%%%%%%%%%%%%%%%%%%%%%%%%%%%%%%%%%%%%%%%%%%%%%%%%%%
\begin{figure}[htbp!]
    \includegraphics[width=7cm]{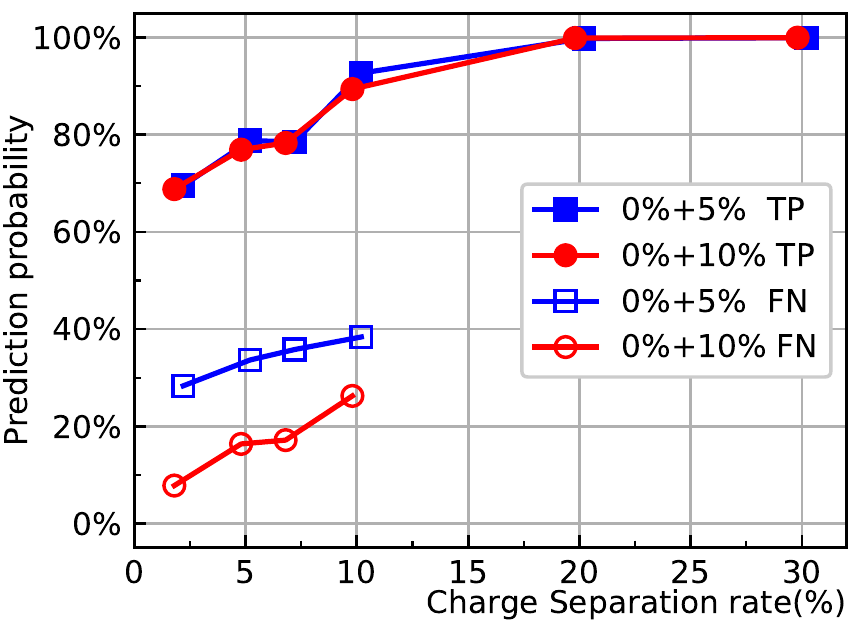}
    \caption{The prediction probability from the networks on data-sets with mixed collision energies and centralities along with varying CS rate.
    }
    \label{fig:MoreCS_test}
\end{figure}
%%%%%%%%%%%%%%%%%%%%%%%%%%%%%%%%%%%%%%%%%%%%%%%%%%%%%%%%%%%%%%%%%%%%%%%%%%%%%%%%%%%%%%%%%%%%%%%%%%
To eliminate the ambiguity of introducing the CME under various conditions, we take two species of data-set with different CS fractions to train the deep CNNs. They are $f = 5\%$ and $10\%$ \enquote{CS} events mixed with equivalent amount of \enquote{no CS} events. The corresponding well-trained CNN models are named as (0\%+5\%) and (0\%+10\%), respectively, in Table \ref{tab:acc}, with their performance on recognizing the CS signal also shown.  The validation accuracy of the (0\%+5\%) model is less than the other one. It indicates that less distinctiveness among the data in (0\%+5\%) case makes it harder for achieving high classification accuracy compared to the more discriminative data in the latter case. In spite of the discrepancy between the two models, their performance is robust against various collision conditions, such as the $\sqrt{s_{NN}}$ and centrality, see Fig.~\ref{fig:Acc_vs_cen} for centrality dependency. The performance of the network also reflects that the CS signals are not totally diminished or contaminated after the collision dynamics, and can be visible with the help of deep learning. As to the generalization ability of the trained network, both models are verified in different collision energies and initial CS setups, which is shown in Suppl.~I. With the increasing of CS fraction $f$, both models demonstrate improvements of the prediction. In Fig.~\ref{fig:MoreCS_test}, predictions are sorted into true-positive (TP) and false-negative (FN) cases, in which true/false is relative to the ground truth label, whereas positive/negative is the predicted label. With CS fraction increasing, the FN curve stops at $f=10\%$ because both models can 100\% correctly recognize CS events with larger CS fraction.

\emph{CME-meter}.-- In effect, the trained neural network is a mapping between the input (charged pions' spectrum) and the output (CS signal), $P(\rho^{\pm}(p_T, \phi))$, which is learned supervisedly from the training data. The output of the network contains 2 components named as ($P_0$, $P_1$) for each input spectrum with $P_0  +  P_1 = 1$. The value of $P_1$ is identified as the probability that the network regards the input spectrum to correspond to the CS class, thus it is intimately related to the CS signal intensity in the event.
%The value $P_1$ is nothing but the intensity of the CS signal given by the neural network, or technically treated as a prediction possibility of the CS signal.
In above, we have demonstrated that the trained network can efficiently decode the initial CS information purely from $\rho^{\pm}(p_T, \phi)$.
In this sense, the network acts as a meter to measure the probability of CME occurring in HICs. In following, we investigate the correlation between $P_1$ and $\gamma$-correlator to reveal a coherent account of this \textit{CME-meter}.

The $\gamma$-correlator measures the event-by-event two-particle azimuthal correlation of charged hadrons, which is considered sensitive to CME~\cite{Voloshin:2004vk}. It is defined as $\gamma_{\rm same} = \langle \cos(\phi_\alpha^{(\pm)}+\phi_\beta^{(\pm)}-2\Phi_R)\rangle$ or $  \gamma_{\rm opp}  = \langle \cos(\phi_\alpha^{(\pm)}+\phi_\beta^{(\mp)}-2\Phi_R)\rangle$ for correlation between same or opposite charges, where $\phi_\alpha^{(\pm)}$ is the azimuthal angle of particle $\alpha$ with positive or negative charge, $\Phi_R$ is the azimuthal angle of the reaction plane ($\Phi_R$ = 0 in this work), and $\langle\cdots\rangle$ represents average over particles in the event. In order to subtract charge-independent backgrounds, $\Delta\gamma = \gamma_{\rm same} - \gamma_{\rm opp}$ is also often used. In order to compare with the CME-meter, we define the relative difference of $\Delta\gamma$ as
\begin{equation}\label{eq:gamma_contrast}
    R_\gamma = \bigg\vert\frac{\langle\Delta\gamma(1)\rangle-\langle\Delta\gamma(0)\rangle}{\langle\Delta\gamma(1)\rangle+\langle\Delta\gamma(0)\rangle}\bigg\vert,
\end{equation}
with 0 and 1 denoting the $\Delta\gamma$ of \enquote{no CS} and \enquote{CS} events as claimed above. The angle bracket means average over events. Similarly, for the trained network, we define
\begin{equation}\label{eq:network_brightness}
    R_{\rm CNN} = \bigg\arrowvert\frac{\langle P_1(1)\rangle-\langle P_1(0)\rangle}{\langle P_1(1)\rangle+\langle P_1(0)\rangle}\bigg\arrowvert,
\end{equation}
where the angle bracket is an average over testing data-set. The relative difference $R_\gamma$ and $R_{\rm CNN}$ can both reflect the difference between \enquote{no CS} and \enquote{CS} classes.

%%%%%%%%%%%%%%%%%%%%%%%%%%%%%%%%%%%%%%%%%%%%%%%%%%%%%%%%%%%%%%%%%%%%%%%%%%%%%%%%%%%%%%%%%%%%%%%%%%
\begin{figure}[htp!]
    \includegraphics[width=8cm]{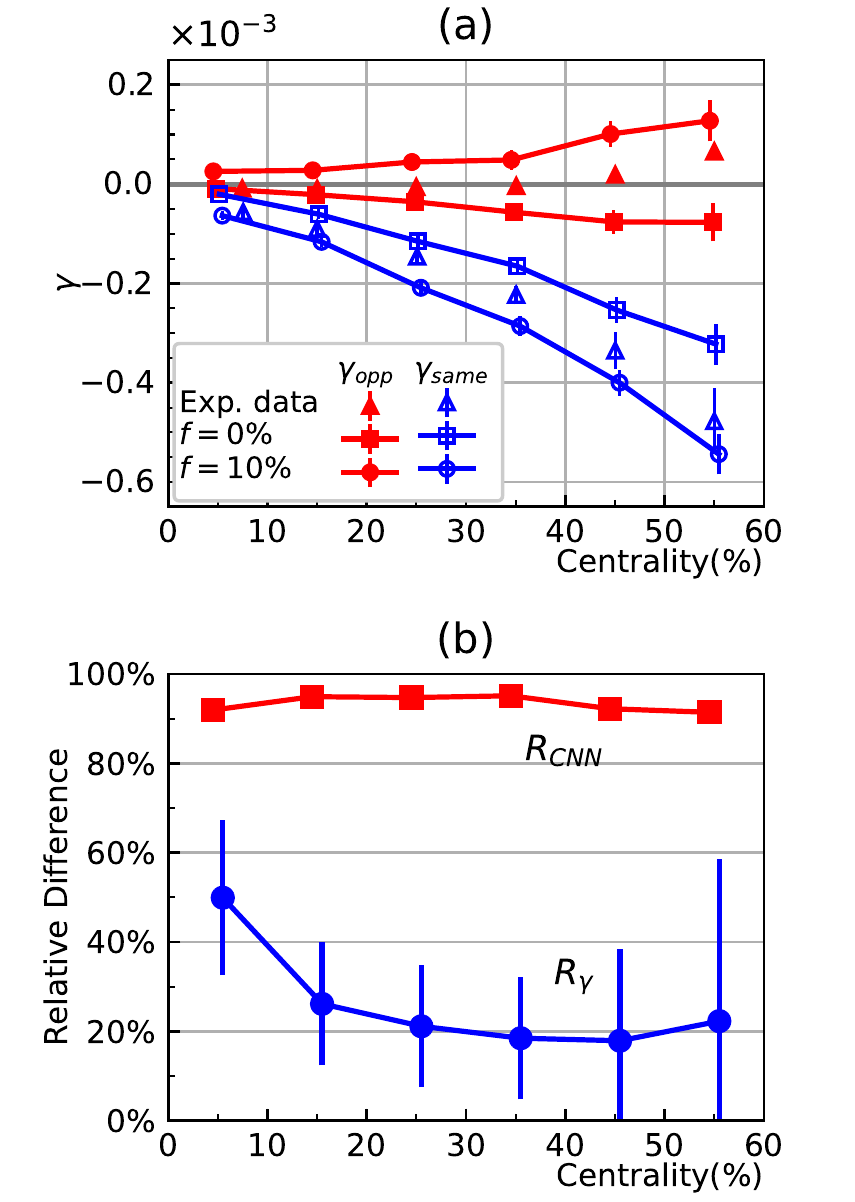}
    \caption{(a) $\gamma$-correlator for events with initial CS ($f=10\%$) and events without CS ($f=0\%$) for Au+Au collisions at 200 GeV. (b) Relative difference of $\Delta\gamma$ and $P_1$ at different centralities.}
    \label{fig:gamma}
\end{figure}
%%%%%%%%%%%%%%%%%%%%%%%%%%%%%%%%%%%%%%%%%%%%%%%%%%%%%%%%%%%%%%%%%%%%%%%%%%%%%%%%%%%%%%%%%%%%%%%%%%
% The detail of Rs' might about to be appended Figure~\ref{fig:gamma}(a) shows the comparison between experimental data and AMPT simulation of event-averaged $\gamma$-correlator for Au+Au collisions at 200 GeV.
In Fig.~\ref{fig:gamma} (a), the $\gamma$-correlator from AMPT simulation for Au+Au collisions at 200 GeV is shown in a centrality range $(0,60\%)$ with experimental data also shown as triangle dots~\cite{starcollaboration:2009azimuthal,starcollaboration:2010observation}. %The square and circle dots are calculated from the AMPT model, which manifests distinct consistency with the experiment.
In Fig.~\ref{fig:gamma} (b), the relative differences, $R_{\gamma}$ and $R_{\rm CNN}$, are plotted at different centralities and demonstrate distinctive comparison. %To compare them fairly, we calculated 100 events average $R_{\gamma}$ which is labeld as $R_{\gamma}^{100}$.%The superscript of $R_\gamma$ denotes the number of events used for computing. Although it is not practical to compute the $R_\gamma^{100}$ in experiments, a generally fair comparison between the $R_{\gamma}$ and $R_{CNN}$ could be provided.
% \blue{Let us only show the case of 100 events to make a fair comparison with CNN.}
The $R_\gamma$ tends to perform worse with increasing centralities, while good and robust performance of $R_{\rm CNN}$ is kept in all centralities. This indicates that possible background contamination (increasing with centrality) may mask $\gamma$-correlator, while does not disturb much the CME-meter. %On the other hand, the remarkable performance of $R_{CNN}$ derived from $P_1$ keeps an effective resolution on CS signals in all centralities.

%%%%%%%%%%%%%%%%%%%%%%%%%%%%%%%%%%%%%%%%%%%%%%%%%%%%%%%%%%%%%%%%%%%%%%%%%%%%%%%%%%%%%%%%%%%%%%%%%%
\begin{figure}[htbp!]
    \includegraphics[width=7cm]{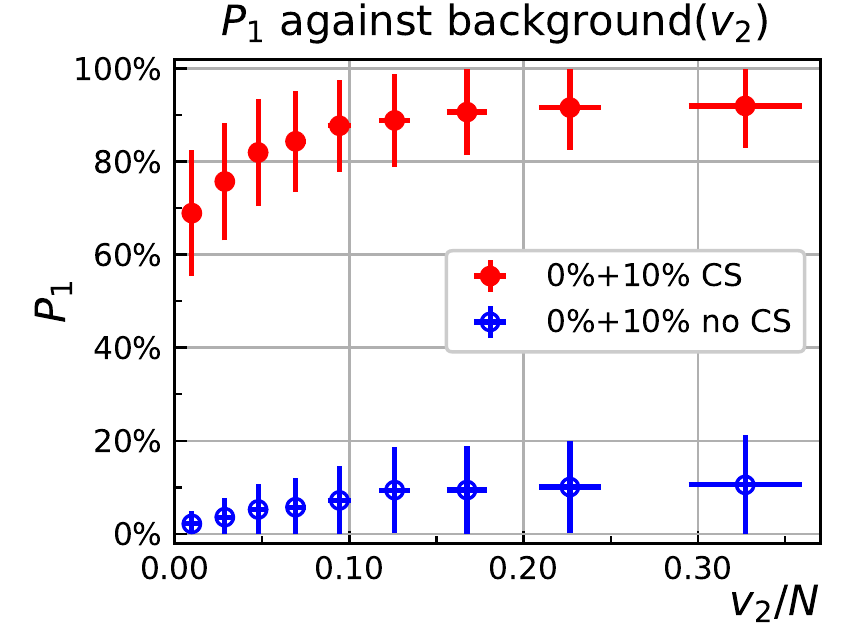}
    \centering
    \caption{The prediction probability from the well-trained network at different $v_2/N$, here $N$ is the multiplicity at mid-rapidity (Au+Au at 200 GeV).}
    \label{fig:v2}
\end{figure}
%%%%%%%%%%%%%%%%%%%%%%%%%%%%%%%%%%%%%%%%%%%%%%%%%%%%%%%%%%%%%%%%%%%%%%%%%%%%%%%%%%%%%%%%%%%%%%%%%%
Such contamination is potentially proportional to the elliptic flow $v_2$. In fact, previous studies reveal that the $v_2$ driven backgrounds can strongly interfere with the CME signal in $\gamma$-corerlator because both the magnetic field and $v_2$ have similar centrality dependence~\cite{Huang:2015oca,Kharzeev:2015znc}. Thus, $v_2$-induced $\Delta\gamma$ can emerge in both \enquote{CS} events and \enquote{no CS} events making $\Delta\gamma$ daunting to distinguish these two classes of events.
%, which leads to the fact that charge separation and elliptic flow are strongly coupled with each other in the $\gamma$-correlator. Thus the $v_2$ is regarded as a background in recognizing CME signals, and it inevitably makes it daunting to isolate CME signals from $v_2$ background. With increasing $v_2$, such interference become worse for CME detection, however,
To examine more closely whether the CME-meter receives $v_2$ influence, we depict $P_1$ versus $v_2/N$ ($N$ is the multiplicity) in Fig.~\ref{fig:v2}. It shows that the deep CNN retains robust to identify the CME signal against the $v_2$ background. Specifically, $P_1$ of \enquote{CS} events changes inconspicuously with $v_2/N$ increasing except for small $v_2/N$, where a slight dependence on $v_2/N$ of $P_1$ is seen. %the background contributed from elliptic flow slightly reduce the value of $P_1$, while at larger $v_2/N$, $P_1$ does not vary with $v_2$ dramatically.
The deep CNN recognizes CS clearly, with the significance to be about $5\sigma$, averaged over all $v_2/N$.
% \blue{why use mid-rapidity $N$?}
%%%%%%%%%%%%%%%%%%%%%%%%%%%%%%%%%%%%%%%%%%%%%%%%%%%%%%%%%%%%%%%%%%%%%%%%%%%%%%%%%%%%%%%%%%%%%%%%%%
\begin{figure}[ht!]
    \centering
    \includegraphics[width=7cm]{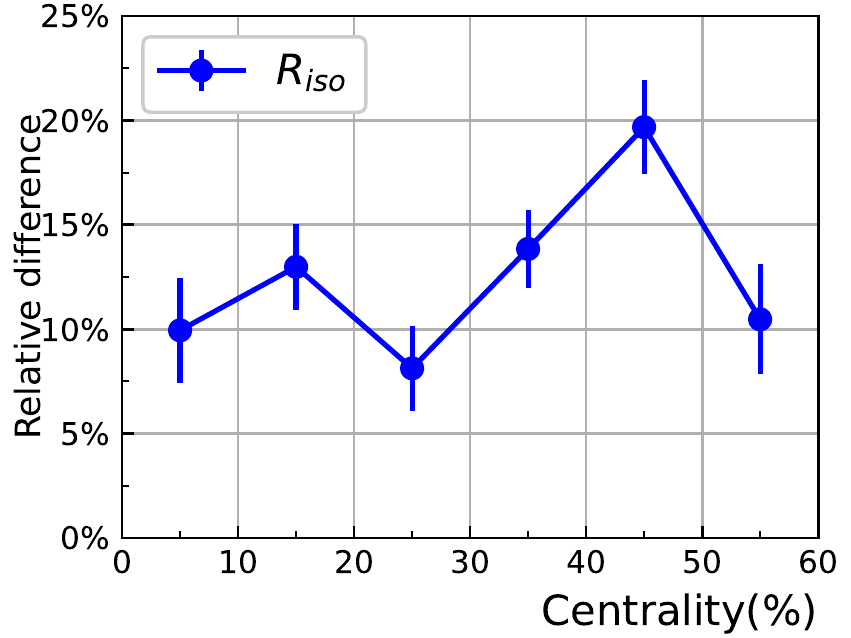}
    \caption{The results of the (0\%+10\%) model on the isobaric collision systems (Ru+Ru and Zr+Zr at 200 GeV).}
    %Since $P_1$ takes value in $[0,1]$, the error bars which exceed $1$ are cut to meet the definition.
    %Tests of the (0\%+5\%) model have also been done, giving even better results. (but hard to show in this figure)}
    \label{fig:isobar}
\end{figure}
%%%%%%%%%%%%%%%%%%%%%%%%%%%%%%%%%%%%%%%%%%%%%%%%%%%%%%%%%%%%%%%%%%%%%%%%%%%%%%%%%%%%%%%%%%%%%%%%%%

The robustness of the CME-meter could also be validated in isobaric collisions~\cite{Voloshin_2010,Deng:2016knn,Deng:2018dut,Xu:2017zcn,Shi:2019wzi}. In our case, the network is trained to recognize the CS signal in Au+Au collisions, whereas its generalization ability is tested by confronting Zr+Zr and Ru+Ru collisions. They are isobaric collision systems that are recently proposed specifically for CME search. Although Zr and Ru are deformed nuclei, moreover, their masses are about half of Au, the CME-meter constructed by the network trained over Au+Au data-set clearly recognized the CME signals in isobaric collision systems, as shown in Fig.~\ref{fig:isobar}, where
%the results shown in Fig.\ref{fig:isobar} make it clear that the \textit{CME-meter} can recognize CME signals in different isobaric collision systems.
the quantity presented is $R_\text{iso}=2(\text{logit}(P_1^\text{Ru})-\text{logit}(P_1^\text{Zr}))/(\text{logit}(P_1^\text{Ru})+\text{logit}(P_1^\text{Zr}))$. It indicates a distinguishable difference between the two isobaric collision systems caused by $P_1^\text{Ru}>P_1^\text{Zr}$ as measured from the CME-meter. Physically, it is because Ru has more protons, which may induce a larger magnetic field and thus cause a larger CS signal in Ru+Ru collisions. The results demonstrate that the CME-meter could offer an alternative way to measure CS signal effectively in a range of collision systems, and it holds the robustness in confronting different test conditions, which is largely due to the joint efforts from a series of prepossessing operations inspired by physical insights including normalization, symmetrization, and boundary condition treatment.

\emph{Interpretable deep learning for CME}.--  The prediction $P_1(\rho^{\pm}(p_T, \phi))$ from the well-trained network could be also understood as a CME-signal response to the spectrum $\rho^{\pm}(p_T, \phi)$, which can be utilized to find the most responsive spectrum via the following variational treatment,
\begin{equation}
    \frac{\delta P_1(\rho^{\pm}(p_T, \phi))}{\delta \rho^{\pm}(p_T, \phi))}=0.
\end{equation}
Specifically, with the pion spectrum to be a variational \textit{Ansatz}, we start from a flat spectrum $\rho^{\pm}(p_T, \phi)={1}/{X}$ with $X=480$ the total number of pixels of the spectrum, which derives $P_1=0$, and gradually tune the functional form of $\rho^{\pm}(p_T, \phi)$ with the variational target to maximize $P_1(\rho_{0}^{\pm}(p_T, \phi))$, that is, to approach $P_1=1$. Note that the trained CME-meter network is fixed, through which gradient of its output with respect to its input, $\delta P_1(\rho^{\pm}(p_T, \phi))/\delta \rho^{\pm}(p_T, \phi))$, can be evaluated via back propagation and is provided as the guidance for the above spectrum tuning. The resultant \enquote{ground state} $\rho_0^{\pm}(p_T, \phi)$ could disclose the crucial patterns manifesting the CS signal in the perspective of the trained network. In machine learning language, the above procedure is the so-called \textit{DeepDream} method~\cite{szegedy:2015going}, in which the variational tuning is implemented as gradient ascent algorithms.
%%%%%%%%%%%%%%%%%%%%%%%%%%%%%%%%%%%%%%%%%%%%%%%%%%%%%%%%%%%%%%%%%%%%%%%%%%%%%%%%%%%%%%%%%%%%%%%%%%%%%%%%%
\begin{figure}[htbp!]
    \centering
    \includegraphics[width=8.4cm]{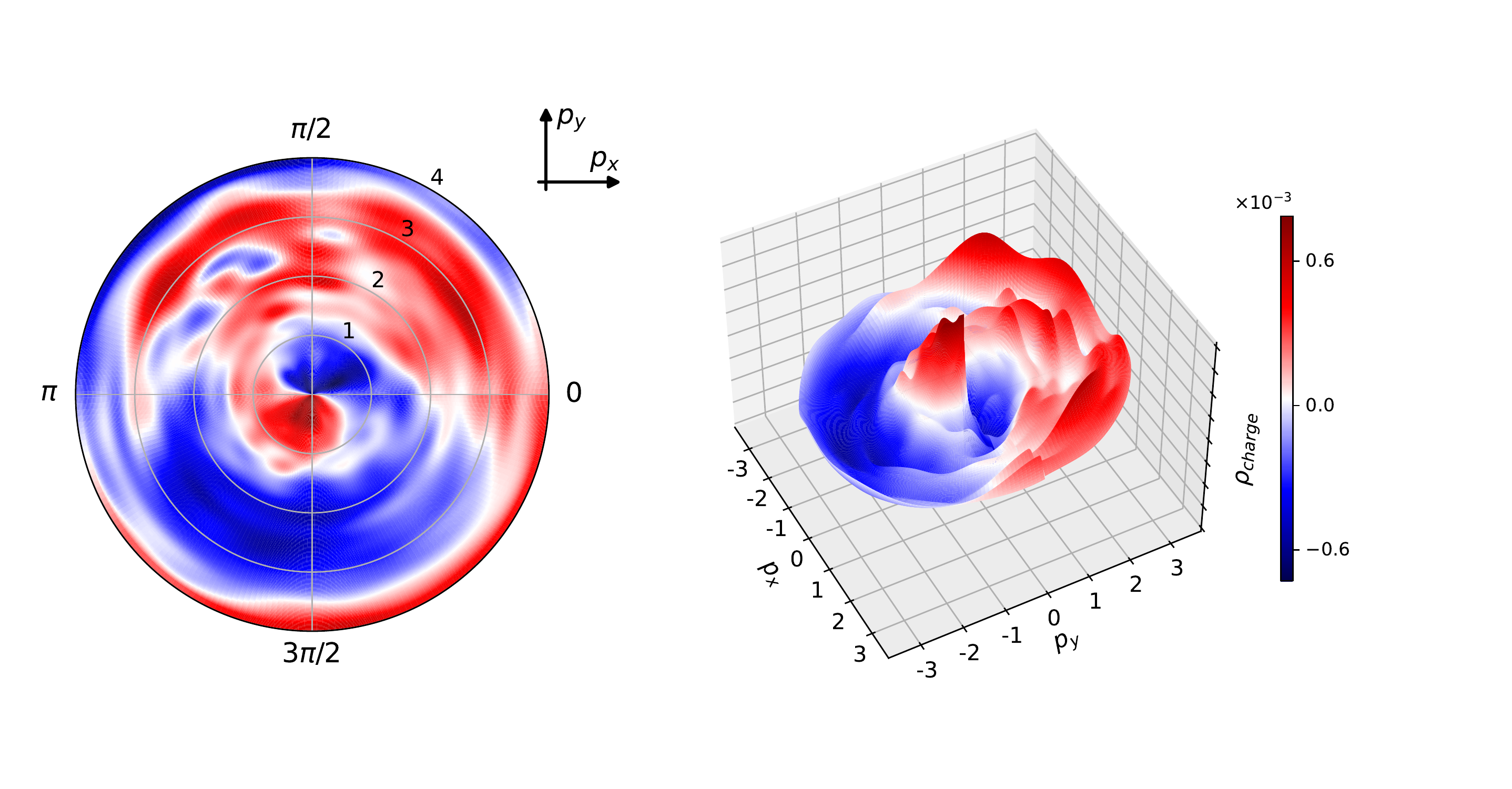}
    \caption{\textit{DeepDream} map for the (0\%+10\%) model.}
    \label{fig:DeepDream}
\end{figure}
%%%%%%%%%%%%%%%%%%%%%%%%%%%%%%%%%%%%%%%%%%%%%%%%%%%%%%%%%%%%%%%%%%%%%%%%%%%%%%%%%%%%%%%%%%%%%%%%%%%%%%%%%

In Fig.~\ref{fig:DeepDream}, the \enquote{ground state} pion spectrum tuned from the \textit{DeepDream} method is visualized (also see Suppl.~II). Although this spectrum may neither be real nor physical, it shows the \enquote{CME pattern} that the network would response most dramatically. Similar visualization method has also been demonstrated in image recognition tasks~\cite{gatys:2016neural}. The $\rho_{0}^{\pm}(p_T, \phi)$ explains the following basic features,
\begin{itemize}
\item \textit{Charge conservation.} During the procedure of variations, the charge conservation is reasonably reserved, which is examined by tracking the charge density of the spectrum.
\item \textit{Dipole structure.} The \enquote{ground state} spectrum from \textit{DeepDream} variation intuitively displays the CS pattern. In the low $p_T$ regime (center), the distribution of pions induces an electric current, or a dipole downward, nevertheless it presents an opposite current which is larger in the high $p_T$ regime ($p_T \sim 3$ GeV).
\end{itemize}
It should be mentioned that $\rho_{0}^{\pm}(p_T, \phi)$ derived from \textit{DeepDream} is a virtual spectrum whose local properties depend on the AMPT simulation and the well-trained neural networks. However, it offers a reliable way to evaluate the effectiveness of $P_1$ in detecting CME and help us reveal the physical contents the machine learns.

\emph{Summary}.-- In this Letter, we propose a deep convolutional neural network (CNN) model to detect CME signal in the simulated data-set from a multiphase transport model. With two different charge separation fractions (5\% and 10\%), the machine is trained to recognize the CME signal under supervision. It is worth noting that this well-trained machine provides a powerful meter to quantify the CME with different collision energies and centralities. The meter also shows a robust performance at different charge separation fractions. In comparison with the conventional $\gamma$-correlator, the CME-meter remains insensitive to the backgrounds dominated by the elliptic flow $v_2$. We also extend the well-trained machine to other collision systems by means of transfer learning. Remarkably, the meter successfully recognize the CME signal and their difference from Zr+Zr and Ru+Ru collisions. It indicates that the knowledge of identifying CME signal in Au+Au collision could be transformed into the knowledge of detecting CME in other collision systems. In the end, \textit{DeepDream}, a method used to visualize the patterns learned by CNNs, is applied as a validation test of adopting $P_1$ to detect the CME. It helps us drill the physical knowledge hidden in the well-trained machine, including charge conservation and special charge distribution.

% Accuracy
% Tests under different energy and centrality bins
% The test of different CS rate f
% Tests on the isobar events
% Deep dream
% summary

%\noindent $^*$ E-mail:
\emph{Acknowledgement}.-- We acknowledge discussions with D. Kharzeev, R. Lacey, and X. N. Wang. The work is supported by NSFC through Grant No.~12075061 and Shanghai NSF through Grant No.~20ZR1404100 (Y. S. Z and X. G. H), by the AI grant at FIAS of SAMSON AG, Frankfurt (L. W. and K. Z.), by the BMBF under the ErUM-Data project (K. Z.),  and by the NVIDIA Corporation with the generous donation of NVIDIA GPU cards for the research (K. Z.).

\bibliography{LearnCME}

\onecolumngrid
\section*{Supplemental Materials}
\subsection*{Test Accuracy}
%%%%%%%%%%%%%%%%%%%%%%%%%%%%%%%%%%%%%%%%%%%%%%%%%%%%%%%%%%%%%%%%%%%%%%%%%%%%%%%%%%%
\begin{figure}[htbp!]
    \includegraphics[width=7.8cm]{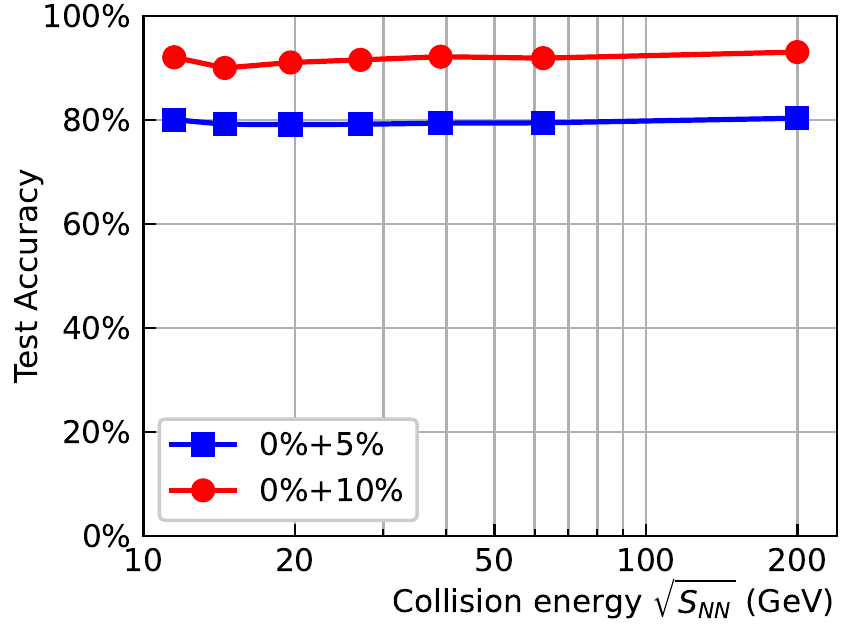}
    \caption{The test accuracy of models on different data-sets with diverse collision energies $\sqrt{s_{NN}}$ which contains mixed centralities.}
    \label{fig:Acc_vs_eng}
\end{figure}
%%%%%%%%%%%%%%%%%%%%%%%%%%%%%%%%%%%%%%%%%%%%%%%%%%%%%%%%%%%%%%%%%%%%%%%%%%%%%%%%%%%
%%%%%%%%%%%%%%%%%%%%%%%%%%%%%%%%%%%%%%%%%%%%%%%%%%%%%%%%%%%%%%%%%%%%%%%%%%%%%%%%%%%
\begin{figure}[htbp!]
    \includegraphics[width=7.8cm]{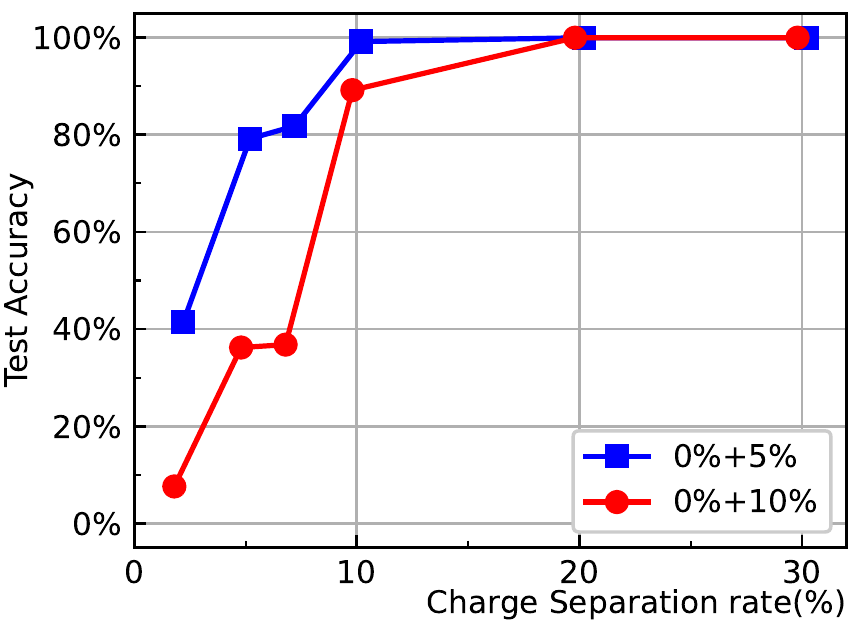}
    \caption{The test accuracy of well-trained models on different data-sets with CS fraction $f$ which contains mixed collision energies and centralities.}
    \label{fig:Acc_vs_CS}
\end{figure}
%%%%%%%%%%%%%%%%%%%%%%%%%%%%%%%%%%%%%%%%%%%%%%%%%%%%%%%%%%%%%%%%%%%%%%%%%%%%%%%%%%%
Although the (0\%+5\%) model does not exceed the (0\%+10\%) model as shown in Fig.~\ref{fig:Acc_vs_eng}, the situation converses in cases with diverse CS fractions. Figure~\ref{fig:Acc_vs_CS}  manifests that the (0\%+5\%) model has a better performance than the (0\%+10\%) one, which behaves as a higher accuracy in testing data-sets. The (0\%+5\%) model could ameliorate the over-fitting problem that caused by modeling the random noise in each event, rather than the intended outputs~\cite{mehta:2019highbias}. It endows the (0\%+5\%) model with a stronger generalization ability. However, (0\%+5\%) model
could capture the CS signal sensitively, which can give rise to the overreaction. It eventually presents a prediction without enough resolution in different cases.

\subsection*{DeepDream}
%%%%%%%%%%%%%%%%%%%%%%%%%%%%%%%%%%%%%%%%%%%%%%%%%%%%%%%%%%%%%%%%%%%%%%%%%%%%%%%%%%%%%%%%%%%%%%%%%%%%%%%%%
\begin{figure}[htbp!]
    \centering
    \includegraphics[width=8.34cm]{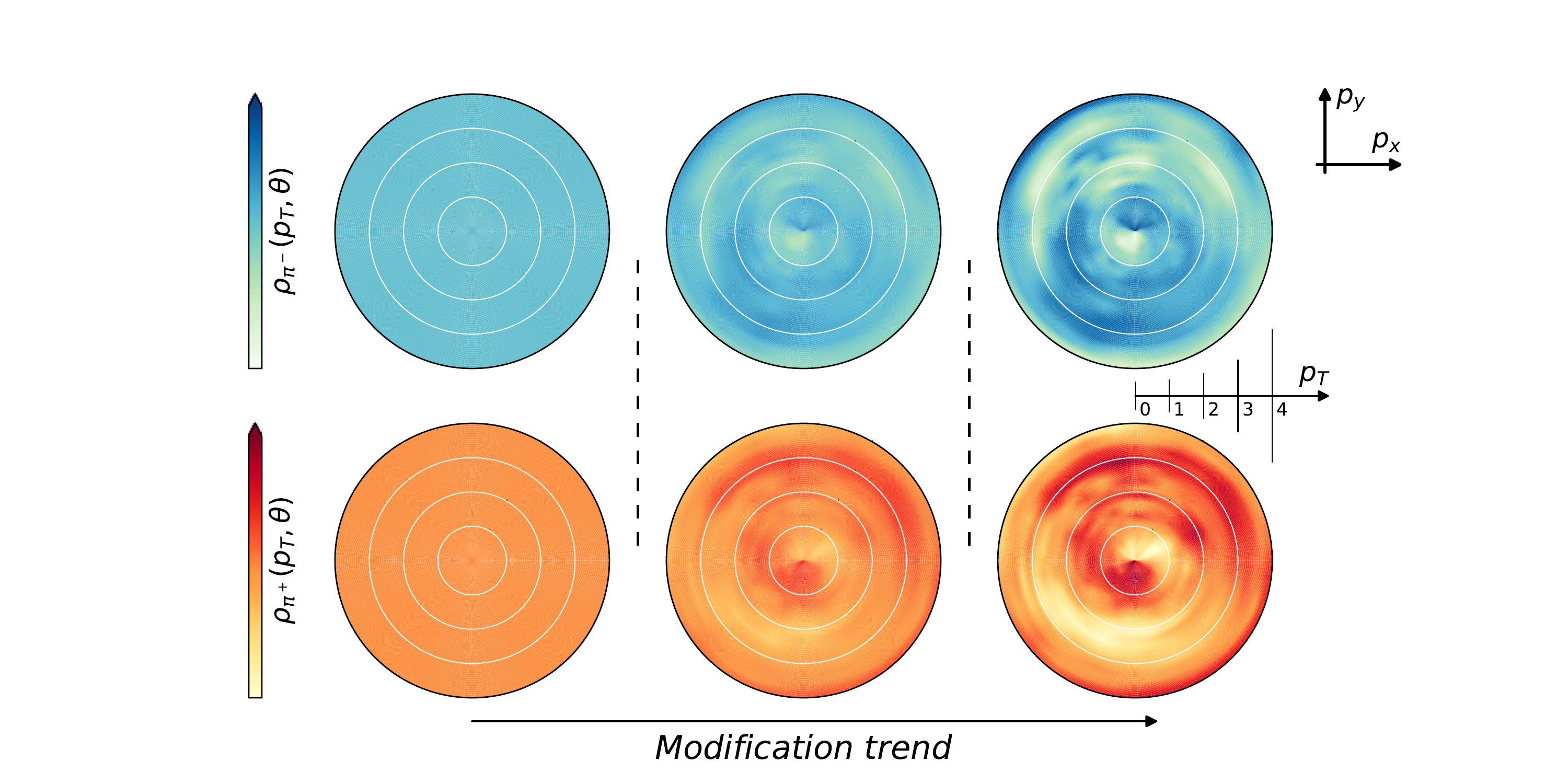}
    \caption{\textit{DeepDream} method using the (0\%+10\%) model.}
    \label{fig:DeepDreamFlow}
\end{figure}
%%%%%%%%%%%%%%%%%%%%%%%%%%%%%%%%%%%%%%%%%%%%%%%%%%%%%%%%%%%%%%%%%%%%%%%%%%%%%%%%%%%%%%%%%%%%%%%%%%%%%%%%%
\textit{DeepDream} modifies an image to increase the activation of certain patterns by gradient ascent method. Figure~\ref{fig:DeepDreamFlow} is the DeepDream result revealing patterns of CS signal. Upper row of polar plots colored blue are spectra of $\pi^-$, and the lower row of polar plots colored orange are $\pi^+$. Two spectra in a column form a sample that we feed and is also fed back by the network. From left to right, \textit{DeepDream} method gradually strengthens the activation of the \enquote{CS} class, thus visualizing the knowledge or physical content learnt in the training.

\end{document}